\documentclass[aps,showkeys,dcolumn,preprintnumbers,notitlepage, secnumarabic,11pt, nofootinbib]{revtex4}
\usepackage{dcolumn}      
\usepackage{bm}           
\usepackage[caption=false]{subfig}

\usepackage{graphicx}
\usepackage{xcolor}
\usepackage{rotating}
\usepackage{amsmath,amssymb}  
\usepackage{bm}  

\usepackage{breqn}
\begin{document}

\title[The dark sector dynamics ]{Interacting Dark Sector: An Isobaric Approximation}
\author{B. Osano}
 \altaffiliation[]{Cosmology and Gravity Group, Department of Mathematics and Applied Mathematics, and \\
 Centre for Higher Education Development,\\ University of Cape Town (UCT), Rondebosch 7701, Cape Town, South Africa.}

 \begin{abstract}
What if the dark sector is not a static entity but rather a dynamic entity with interactive components of the universe? This intriguing hypothesis raises important questions regarding the phenomenological behaviour of such an evolving system. In this study, we explore the simultaneous evolution and interaction of two hypothetical components within the context of an interacting dark sector, examining their implications for our understanding of cosmological dynamics.\end{abstract}

\maketitle
\section{\label{intro}Introduction}  Standard cosmological modelling relies on a dynamical framework that requires the characterisation of the material and energy content of the Universe in some form. This framework is typically encapsulated by several key quantities: the average 4-velocity \( u^{\mu} \), the mass-energy density \( \rho \), the pressure \( p \), and the energy-momentum tensor \( T^{\mu\nu} \). For a perfect fluid, the energy-momentum tensor is expressed as:

\begin{eqnarray}
T^{\mu\nu} = (\rho + p) u^{\mu} u^{\nu} - p g^{\mu\nu}
\end{eqnarray}
where \( g^{\mu\nu} \) denotes the metric tensor of spacetime. This formulation, which captures the essential properties of a perfect fluid, allows for a comprehensive description of the dynamical evolution of cosmic structures under the influence of gravitational interactions. The 4-velocity \( u^{\mu} \) characterises the flow of the fluid, while \(\rho \) and \(p \) denote the energy density and pressure, respectively. These quantities are crucial in determining the expansion and curvature of the Universe as described by the Einstein field equations. The energy conservation is given by \(\nabla_{\nu}T^{\mu\nu}=0\) subject to various energy conditions: the null energy condition, which requires \( \rho + p \geq 0 \); the weak energy condition, which stipulates \( \rho \geq 0 \) and \( \rho + p \geq 0 \); and the strong energy condition, which mandates \( \rho + 3p \geq 0 \) and \( \rho + p \geq 0 \). Together, these conditions seemed to ensure a physically viable description of cosmological dynamics until the discovery of cosmic acceleration \cite{Perl, Riess}. For such acceleration to occur within this framework, the strong energy condition has to be violated, as we will reiterate in our review of the equations of motion in section (\ref{fEq}).

In this context, the fundamental dynamical description necessitates the incorporation of an equation of state. A commonly utilised form is that of a barotropic fluid, characterised by the relation \( p = \omega \rho \), where \( \omega \) is a dimensionless parameter that encapsulates the properties of the fluid. This basic formulation has proven to be an invaluable tool in phenomenological studies, facilitating a deeper understanding of the processes governing the evolution of the Universe.

Evidence for the evolution of dark energy is increasingly supported by various analyses that integrate data from the Dark Energy Spectroscopic Instrument (DESI) alongside other observational datasets \cite{Desi}. Before the discovery of cosmological acceleration \cite{Perl, Riess}, cosmologists largely accepted the premise that pressure is non-negative. Consequently, they characterised a barotropic fluid's pressure using the equation of state \(p=\omega\rho\), where \(\rho\) denotes the energy density. However, the acceptance of cosmological models featuring negative pressure has since emerged, opening avenues for consideration of materials with a parameter \(\omega < 0\). 

As established in the analysis of the Friedmann equations, the condition for cosmic acceleration necessitates that \(\omega < -1/3\). The possibility of evolving dark energy prompts a reevaluation of the foundational principles governing barotropic cosmological fluids. There are at least three potential approaches to advance this inquiry. Several possible equations of state have been constructed and subjected to different observational datasets. Let us denote pressure as a function of energy density: \(p = f(\rho)\). The following strategies can be considered:
\begin{itemize}
    \item Retain the barotropic form while utilising various datasets and their combinations to precisely determine the value of \(\omega\), for example, in \cite{Desi}.
    \item Maintain the barotropic nature of the fluid but allow \(\omega\) to evolve. The effective \(\omega\) can then be constructed properly. We will list examples of these shortly. 
    \item Reconsider thermodynamics in the formulation of the equations of motion\cite{Prigogine:1989zz}. 
    \item Abandon the barotropic fluid approximation.
\end{itemize}
Each of these approaches has the potential to enhance our understanding of dark energy and its implications for the universe's dynamics. The first strategy, which gained prominence following the discovery of cosmic acceleration, remains a focal point of ongoing research. Although efforts to accurately determine cosmological parameters have been bolstered by advancements in technology and data analysis methodologies, the complexities and challenges inherent in these studies have given rise to the second strategy, which is the primary focus of this letter. To begin, we will examine some of the equations of state associated with dark energy that have emerged in the research literature concerning evolving equations of state.
\begin{table}[h]
\caption{Examples of dark energy EoS} 
\centering 
\begin{tabular}{l ll} 
\hline\hline 
$ \omega(z)=\omega_{0}+\omega_{1}z$ &\cite{Linder}& \\ 
$ \omega(z)\equiv\omega_{Q}(z)=-1+\frac{1+z}{3}$ &\cite{Ger}& \\ 
$ \omega=\frac{2q-1}{3}$ &\cite{Avil}&\\ 
$ \omega_{X}=-\frac{1}{3}(1+\frac{\Omega_{m,0}}{\Omega_{X,0}})$&\cite{Chev}&\\ 
$ \omega(z)=\alpha(1-\frac{\rho_{0}}{\rho})$&\cite{Bab24}&\\ 
$ \omega(z)=\omega_{0}+\omega_{a}\frac{z}{(1+z)}$&\cite{CPL,Linder}&\\ 
$ \omega(z)=\omega_{0}+\omega_{a}\frac{z}{(1+z)^{p}}, p=1, 2$&\cite{Jass}&\\ 
$ \omega(z)=\omega_{0}+\omega_{1}\frac{z(1+z)}{(1+z^2)}$&\cite{Bar}&\\ 
$\omega_{z}=\omega_{eff:DE}(z)-\frac{Q}{3H\rho_{DE}}$& \cite{Esca}&\\ 
$\omega(z)=-1+(1+\omega_{DDM}(z))X+(1+\omega_{DDE}(z)Y)$&\cite{bob}&\\ 
$\omega_{\varphi}=\frac{1+\omega_{n}}{1+(a_{c}/a)^{3(1+\omega_{n})}}$&\cite{Poulin}&\\ 
\hline 
\end{tabular}
\label{tab:hresult}
\end{table}
 
\section{Phase Transition}
To contextualise our study, we must first consider cosmological phase transitions. Phase transitions are fundamentally associated with symmetry breaking, resulting in alterations to the state of matter. The symmetry breaking originates from the ground state rather than from the laws of motion, even though these equations are fundamental to our phenomenological analysis. It is widely believed that at least two significant phase transitions occurred in the early universe. The first involves the transition from a quark-gluon plasma to a hadron gas, commonly referred to as the QCD transition \cite{Rischke,Rajagopal}. 
The second pertains to spontaneous symmetry breaking, which facilitates the acquisition of gauge-invariant masses by particles within the Standard Model \cite{Higgs,Anderson}. Moreover, it is conceivable that additional phase transitions could take place; for instance, a strong first-order transition may be capable of generating dark matter \cite{Seto}. Thus, it is prudent to acknowledge that our understanding of potential phase transitions is far from exhaustive. This letter advocates for a significant reorientation from the conventional formulation in cosmological studies, positing that a varying equation of state should be regarded as the fundamental parameter in the ensuing dynamical analysis of the dark sector. We imagine the existence of substances that continuously undergo phase transitions \cite{Mazum}, a phenomenon that deviates from conventional principles in standard physics or standard models. Models extending beyond the Standard Model have been explored in the context of a confined dark sector, yielding intriguing implications for dark matter and gravitational waves \cite{Pedro}. These investigations suggest that the interacting dark sector merits deeper exploration, especially regarding its potential interactions and implications for cosmological phenomena. Notably, the early dark energy (EDE) model was found to be dynamically significant at redshifts \( z \gg 1 \) and thought to have reached an equation of state \( \omega \approx -1 \) at certain stages of its evolution. This observation prompts us to consider: could the early dark energy be just one manifestation of a much broader class of dark energy models, some of which may undergo phase transitions in the later epochs of the universe?\cite{Elor} On the other hand, evolving dark energy (EDM) has recently been studied \cite{Chen}, where an explicit example is given. What we see is a dark sector whose constituents interact and manifest evolving equations of state. 
This proposition necessitates a thorough examination of the thermodynamic properties associated with such a substance; however, we will defer that analysis to future studies. Instead, we commence this inquiry with the assumption that such a substance exists. Consequently, our approach aims to deepen our understanding of the universe's fundamental dynamics by incorporating the variability of the equation of state into the theoretical framework. From the above list of proposed equations of state, it appears that they can be categorised into two classes. These are the linear equation of state - $\omega'=constant$ and the non-linear equation of state $\omega'\ne constant$, where the derivatives are with respect to the redshift. In addition to the two classes is the constant equation of state case (iii) $\omega'=0$ ( non-evolving equation of state). We focus on the scenario in which the equation of state parameter, denoted as \(\omega\), is non-linear. The ideal measure of change would be the intervals between transitions between different states. However, in our current study, we lack direct measurements of these intervals. Instead, we will examine a scenario where this temporal variation is monitored through redshift measurements, utilising a phenomenological approach to analyse the resultant system of equations.\section{\label{fEq}Friedmann Equations/Equations of motion}
The equations of motion for the $\Lambda$CDM takes the form \begin{eqnarray}
\left(\frac{\dot{a}}{a}\right)^{2}&=& \frac{8\pi G}{3}\sum_{i=A,B,C}\rho_{i} - \frac{k}{2a^2} + \frac{\Lambda}{3}\\
\frac{\ddot{a}}{a} &=& -\frac{4\pi G}{3}\sum_{i=A,B,C}(\rho_{i}  + 3p_{i} ) + \frac{\Lambda}{3},
\end{eqnarray} where $a$ is the scale-factor, $G$ is the Newtonian gravitational constant, and $\Lambda$ is the cosmological constant. We will adopt the convention, $8\pi G=1$, and assume a flat geometry $k=0$. To further aid discussions and to link to observation, we will also convert and work in the logarithmic time scale, where prime $'=d/d\eta$ such that $\eta=\log a$. The Friedmann equations in terms of the log scale, $'$, assume the form:
\begin{eqnarray}
1=\left(\frac{{a'}}{a}\right)^2&=& \frac{8\pi G}{3H^2}\sum_{i=A,B,C}\rho_{i} - \frac{k}{(aH)^2} + \frac{\Lambda}{3H^2},\\
\frac{{a}''}{a} +\frac{H'a'}{Ha}&=& -\frac{4\pi G}{3H^2}\sum_{i=A,B,C}(\rho_{i}  + 3p_{i} ) + \frac{\Lambda}{3H^2},
\end{eqnarray} 

While the presentation may lack the elegance of a proper time scale, it will facilitate the discussion regarding redshift. It is important to note that one can operate within the framework of proper time and convert to a logarithmic time scale as necessary
\section{Density evolution Equations for barotropic fluids}
We consider a system comprising three fluid species, denoted as \( A \), \( B \), and \( C \). The species \( A \) and \( B \) are characterised by a barotropic equation of state and are permitted to interact with one another. In contrast, species \( C \) is non-interacting with the first two species and is treated as a dust-like component within our framework. The evolution equations governing the dynamics of these three species take the following form:
\begin{eqnarray}\label{eq1}
{\rho}'_{A}&=&-3(1+\omega_{A})\rho_{A}+\frac{Q_{AB}}{H}\\\label{rhodots2}
{\rho}'_{B}&=&-3 (1+\omega_{B})\rho_{B}+\frac{Q_{BA}}{H}\\\label{rhodots3}
{\rho}'_{C}&=&0,
\end{eqnarray} where \(Q_{BA}=-Q_{AB}\) is the interaction term and involves species $A$ and $B$. We will give a detailed discussion of the interaction in section (\ref{interact}). For now, we take this coupled system of equations to be closed.
\section{Isobaric approximation}
We commence our analysis with barotropic fluids and proceed to articulate the equation of state (EoS) in terms of both pressure and density. Our approach is predicated on the isobaric approximation for interacting fluids. We consider two distinct fluids, which we denote as \( A \) and \( B \). The EoS for these fluids is represented by \( \omega_A \) and \( \omega_B \), respectively, along with their corresponding pressures and densities. In our examination, we focus on the case of a dynamical EoS, permitting these parameters to evolve. We consider the general case in which the pressure \( p \) is a function of the density \( \rho \), expressed as \( p = f(\rho) \). The phenomenological direct proportionality between pressure and density enables us to define a class of barotropic fluids characterised by an equation of state parameter \( \omega \), such that \( p = \omega \rho \). In standard cosmological modelling, the parameter \( \omega \) is typically fixed for a given fluid species, although both pressure and density evolve. In isolation, a fluid species characterised by a constant equation of state would be said to evolve in a manner that maintains the constancy of its equation of state parameter. However, within the framework of our analysis, we seek to relax this constraint, permitting the equation of state to vary dynamically as the system evolves. In this context, the equation of state is treated as a function of the scale factor, which is intrinsically linked to the cosmological redshift. This approach allows us to explore the implications of a time-varying equation of state on the dynamics of the fluid and the broader cosmological evolution. We observe that the equation of state parameter, denoted as \(\omega(z)\), is defined by the ratio of pressure \(p(z)\) to energy density \(\rho(z)\):

\begin{equation}
\omega(z) = \frac{p(z)}{\rho(z)}.
\end{equation}

In this framework, the evolution of both pressure and density contributes to the corresponding evolution of the equation of state. Assuming the interacting species model of the universe as previously described, we consider a hypothetical scenario in which the universe transitions into a period characterised by constant pressure, defined by the redshifts \( z_{\text{in}} \) and \( z_{\text{out}} \). During this interval, we can reformulate the dynamics of the universe in terms of the equations of state corresponding to the competing species present.

Specifically, we can represent this evolution mathematically by expressing the equations of state for each species involved in the interaction. This formulation allows us to analyse how the dynamics of the universe can be understood through the interplay of these competing equations of state, thereby providing insight into the influence of each species on the overall cosmological behaviour during this period of constant pressure.

Let us denote the equations of state for the interacting species as \( \omega_A(z) \) and \( \omega_B(z) \), where each describes the relationship between pressure and energy density for the respective species. The evolution of the total equation of state during this interval can then be characterised by the weighted contributions from each species, reflecting their respective dynamics and interactions.
We can mathematically express the evolution of the universe during the period of constant pressure through the equations of state of the competing species, enabling a comprehensive examination of the underlying physical processes at play.
\begin{equation}
\omega'(z) = \frac{p'(z)}{\rho(z)} - \omega(z) \frac{\rho'(z)}{\rho(z)}.
\end{equation}

In our analysis, we examine the instantaneous isobaric condition, where the pressure derivative \(p'(z)\) is zero. Under this assumption, the equation simplifies to:

\begin{equation}\label{eqw}
\omega'(z) = -\omega(z) \frac{\rho'(z)}{\rho(z)}.
\end{equation}

This formulation allows us to investigate the relationship between the dynamics of the equation of state and the evolution of the energy density in the context of a fluid that maintains constant pressure for a prescribed period.

\section{Evolution equations for equations of state}
By applying the logarithmic derivative to equation (\ref{eqw}) for each species and utilising the system described in equation (\ref{eq1}), we obtain the following results:
\begin{eqnarray}\label{s1eq1}
\omega_{A}'&=&3 (1+\omega_{A})\omega_{A}-\left(\frac{Q_{AB}}{H\rho_{A}}\right)\omega_{A}\\\label{s1eq2}
\omega_{B}'&=&3 (1+\omega_{B})\omega_{B}+\left(\frac{Q_{AB}}{H\rho_{B}}\right)\omega_{B}.
\end{eqnarray} It is noteworthy that the interaction term is the critical component that maintains the coupling between the evolution of the two species. In its absence, the equations of state (EoS) for each species would evolve independently. This observation underscores the necessity of a comprehensive understanding of the role that interactions play in the overall dynamics of the system.
\section{\label{interact}The Interaction or coupling term}
We should not anticipate any additional equalities between the last terms in the system of equations (\ref{s1eq1}-\ref{s1eq2}) unless we impose such conditions artificially. Any such imposition must be justified by physical reasoning if it is to be considered valid or meaningful within the context of the underlying theory. Initially, it is important to recognise that the interaction term must be proportional to the densities, based on the evolution equations governing those densities. We consider two different ansatz in this analysis  
\subsection{\label{q1} Type I: \(Q_{AB}=\alpha\omega_{B}\rho_{B}H\)}
 In this case, $\alpha=\alpha(z)$ may be thought of as the state-altering proportionality term. Substituting this into equations (\ref{s1eq1}) and (\ref{s1eq2}) yields:
 \begin{eqnarray}
 \omega'_{A}&=&3 (1+\omega_{A})\omega_{A}-\frac{\alpha}{\delta}\omega_{B}\omega_{A}\\
 \omega'_{B}&=&3 (1+\omega_{B})\omega_{B}+\alpha\omega_{B}^2,
\end{eqnarray} where $\delta=\rho_{A}/\rho_{B}$. Before proceeding with the analysis of this system of equations, it is essential to clarify two key terms. The parameter \(\alpha\), which is a function of redshift, must satisfy the condition \(\alpha(z) \to 0\) as \(z \to 0\). This condition is critical as it ensures that the interaction term approaches zero in the late-time limit. A plausible functional form for \(\alpha(z)\) that meets this requirement is \(\alpha(z) = e^{-z}\). This formulation effectively captures the desired asymptotic behaviour of \(\alpha\) in the context of redshift. The second term that needs clarification is $\delta$. Both $\rho_{A}$ and $\rho_{B}$ are functions of the redshift. The implications of this observation indicate that the ratio is not a constant value; rather, it varies in accordance with the changing densities of the respective components. In our investigation of the transition between the dominance of one species and another, we will focus on the ratio defined within the bounds \(0 \leq \delta \leq 1\). 

There are two primary methodologies for characterising the parameter \(\delta\). The first approach entails utilising observational data to derive an appropriate ratio that reflects the current state of the system. This method appears to be the most logical, as it grounds our analysis in empirical evidence. Alternatively, we could employ a smoothing function of the variable \(z\) that adheres to the specified requirements for \(\delta\). Using the current estimates of density of dark matter and dark energy,\(\rho_{A}=0.68\) and \(\rho_{B}=0.27\),
\(\delta\approx 2.52\).

In this study, we will adopt the first approach, applying this formalism to examine the onset of late-time cosmic acceleration. In this context, we will approximate the densities of the two competing components: dark matter and dark energy. By doing so, we aim to gain an understanding of the dynamics governing the transition between these influential forces in the late-time evolution of the universe. 
\begin{figure}[htbp]
\begin{center}\caption{$\omega_{A}$ vs $\omega_{B}$, $\alpha=e^{-\eta}$ and $\delta\approx2.52$. The fixed points for the system are : $(\omega_{A}^{*},\omega_{B}^{*}) =(-0.1, 0),(0,-0.75)$ and ($0,0)$.}
\includegraphics[width=0.95\columnwidth]{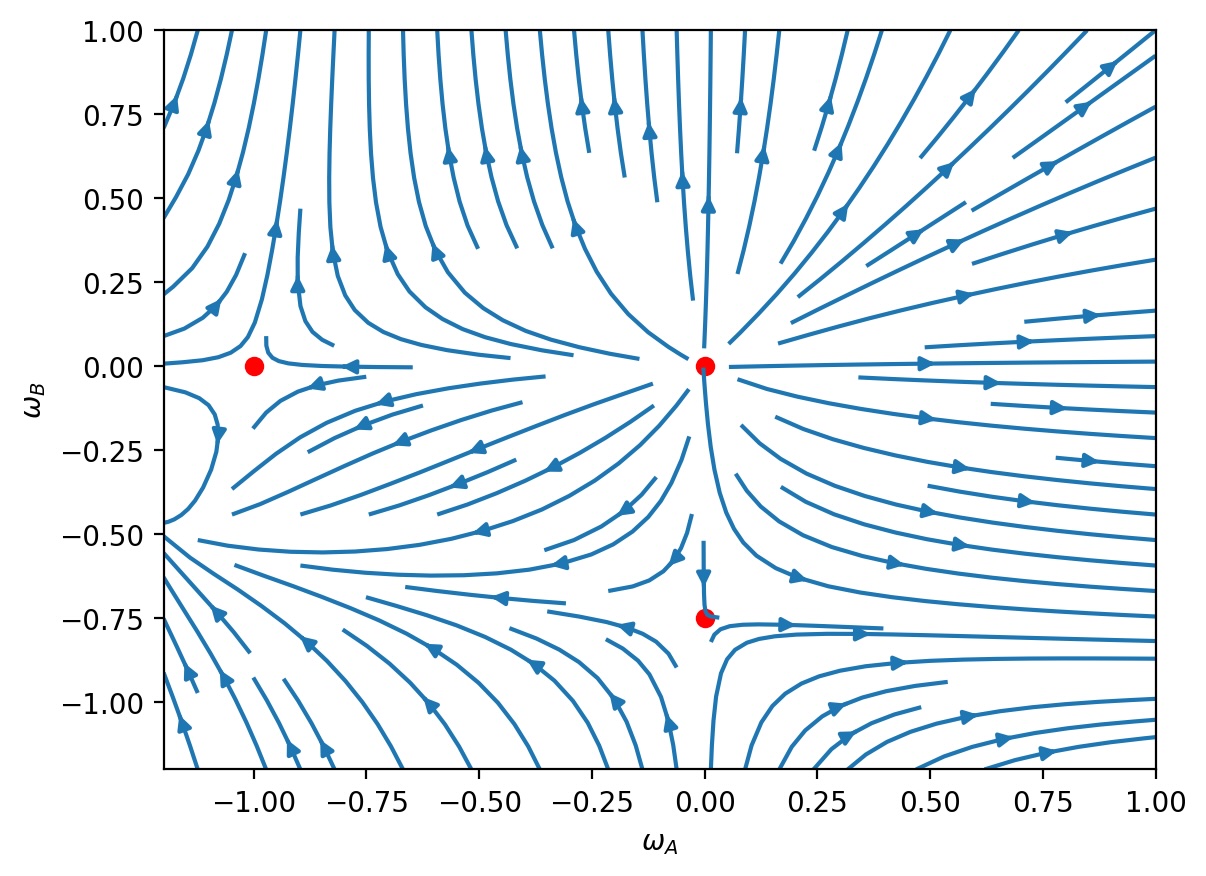}%
\label{fig1}
\end{center}
\end{figure}
The phase portrait of the system, as determined by the proposed ansatz for \(\alpha\) and the fixed ratio \(\delta\), reveals two unstable fixed points located at \((-1, 0)\) and the origin, \((0, 0)\). The implications of these fixed points, along with their respective stability characteristics, will be examined in detail in subsequent sections.

\subsection{\label{q2} Type II: \(Q_{AB} = (\alpha\omega_{a} \rho_{A} + \beta \omega_{B}\rho_{B}) H. \) }
The system in this case takes the form:

\begin{eqnarray}\omega'_{A}&=&3 (1+\omega_{A})\omega_{A}-(\alpha\omega_{A}+\frac{\beta}{\delta}\omega_{B})\omega_{A}\\\omega'_{B}&=&3 (1+\omega_{B})\omega_{B}+(\alpha\delta\omega_{A}+\beta\omega_{B})\omega_{B},
\end{eqnarray} wher again, \(\delta=\rho_{A}/\rho_{B}\).
To analyse the system. We can, as we did in the previous section, set \(\alpha=e^{-\eta}\) and \(\delta=2.52\). If we allow \(\beta\) to scale like \(\alpha\) then the phase portrait take the form:
\begin{figure}[htbp]
\begin{center}\caption{$\omega_{A}$ vs $\omega_{B}$, $\beta=\alpha=e^{-\eta}$ and $\delta\approx2.52$. The fixed points for the system are : $(\omega_{A}^{*},\omega_{B}^{*}) = (0,0)$ and $(0,-0.75)$}
\includegraphics[width=0.95\columnwidth]{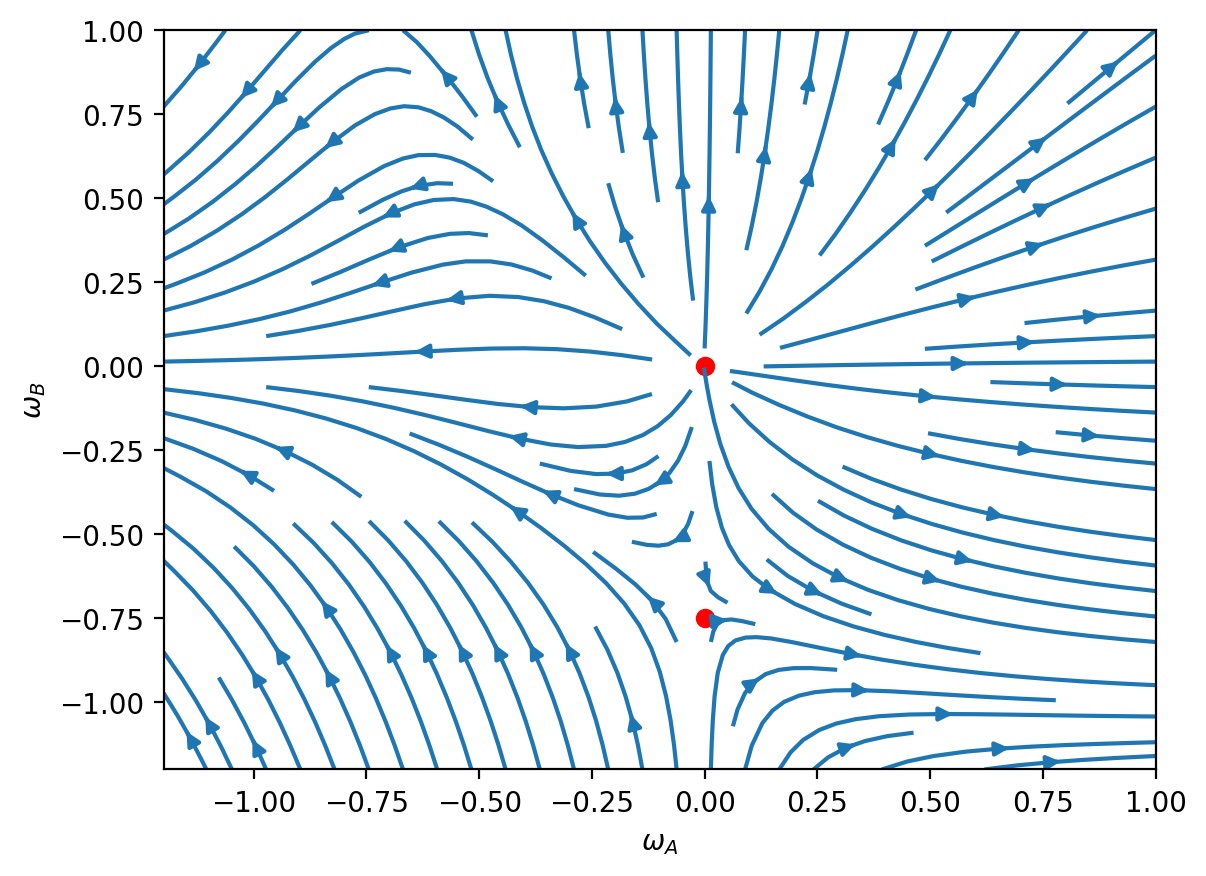}%
\label{fig3}
\end{center}
\end{figure}

\subsection{Type III: $Q_{AB}=(\alpha\omega_{A}\rho_{A}+\beta\omega_{B}\rho_{B})H$ with varying $\delta=\rho_{A}/\rho_{B}$}  
\begin{figure}[htbp]
\begin{center}\caption{$\omega_{A}$ vs $\omega_{B}$, $\delta=\alpha=+e^{-\eta}$ . The fixed points for the system are : $(\omega_{A}^{*},\omega_{B}^{*}) =(-1,0), (-1, -1)$, ($0,0)$ and $(0,-0.75)$}
\includegraphics[width=0.85\columnwidth]{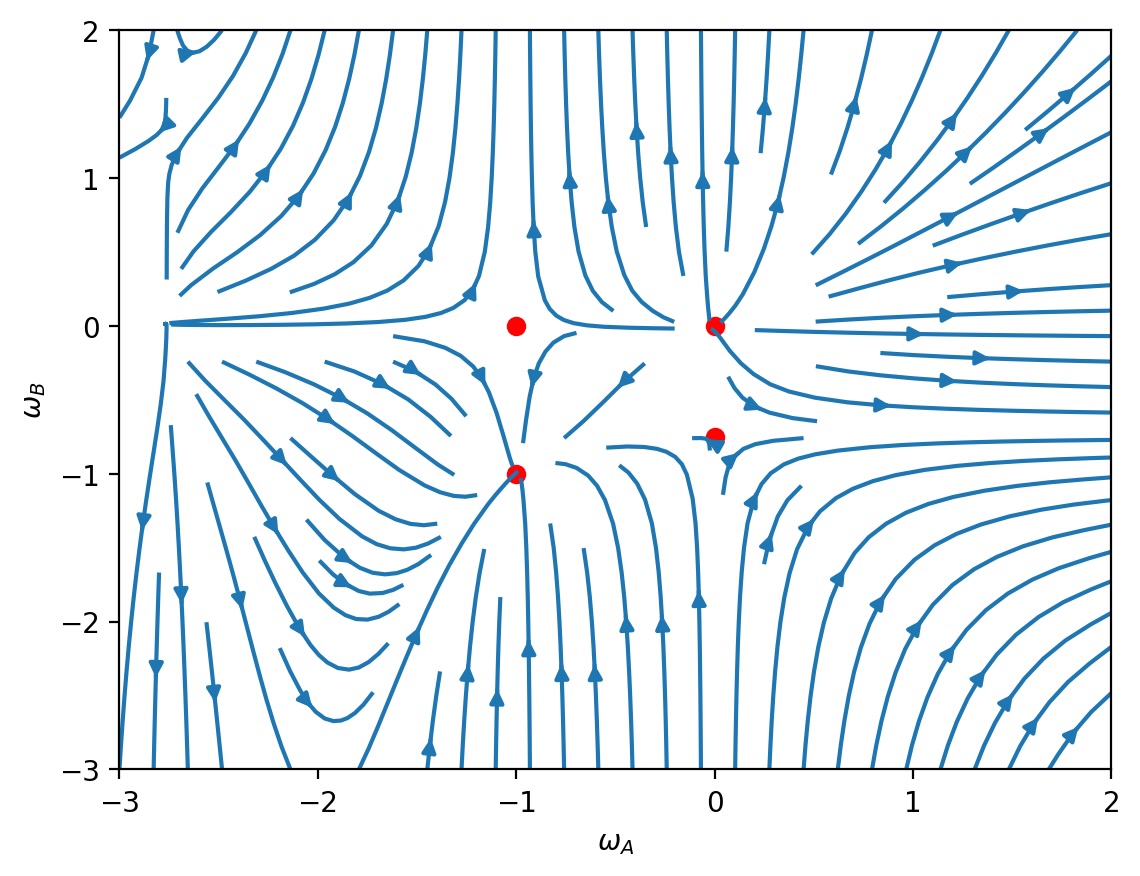}%
\label{fig5}
\end{center}
\end{figure}

From this, we can derive the time derivative of \(\delta\) as follows:
\begin{eqnarray}
\delta' = \frac{\rho'_{A}}{\rho_{B}} - \delta \frac{\rho'_{B}}{\rho_{B}}.
\end{eqnarray} It can be readily demonstrated that this expression simplifies to:

\begin{eqnarray}
\delta' = 3(\omega_{B} - \omega_{A})\delta + (\alpha + \beta)\delta + \beta + \alpha\delta^2.
\end{eqnarray}
In this case, the closed system of equations takes the form:
\begin{eqnarray}\omega'_{A}&=&3 (1+\omega_{A})\omega_{A}-(\omega_{A}\alpha+\frac{\beta}{\delta}\omega_{B})\omega_{A}\\\omega'_{B}&=&3 (1+\omega_{B})\omega_{B}+(\alpha\delta\omega_{A}+\beta\omega_{B})\omega_{B}\\
\delta'&=&(3\omega_{B}-3\omega_{A} +\alpha+\beta)\delta+\beta+\alpha\delta^2.\end{eqnarray} This system, whose phase-potrait we present in FIG \ref{fig5}, is more complex than the previous two. 

\section{Discussion and Conclusion}
The nature of dark energy remains one of the most profound mysteries in contemporary cosmology, with observational data suggesting that its equation of state parameter, \(\omega\), is approximately \(-1\), yet potentially variable\cite{Desi}. In this study, we have taken an innovative approach by investigating substances undergoing multiple phase transitions, thereby allowing for a dynamic equation of state. Our exploration is built upon several assumptions:
\begin{itemize}
\item[i.] We posited that interaction within the dark sector can be quantified through the term \(Q_{AB}\) in our equations of motion.
   
\item[ii.]We permitted the equations of state of the interacting components to evolve as functions of redshift, \(\omega(z)\). This approach represents a more extreme case, where phase transitions are considered to occur on the cosmological timescale of redshift. It is essential to note that this hypothesis is indeed an extreme manifestation of our theoretical framework.

\item[iii.] We adopted the isobaric approximation, asserting that during specific intervals, the pressure remains constant while allowing for the interplay between density and the equation of state.
\end{itemize}
Our investigation of three distinct ansatz for the interaction term reveals that all fixed points are unstable except for one. In particular, the fixed point \((\omega^{*}_{A},\omega^{*}_{B})=(-1,0)\), which be interpreted as representative of $\omega_{DE}=-1$ ( evolving Dark energy) and $\omega_{DM}=0$ ( evolving Dark matter) respectively is unstable and suggestive of a Big Rip scenario. The only stable fixed point is at \((\omega^{*}_{A},\omega^{*}_{B})=(-1,-1)\) within Type III. 
This finding is particularly intriguing, as it suggests that both species A and B may interact in a self-consistent manner, and that appears as interaction between two different species is just self-interaction of the same species - an evolving and self-interacting dark energy. Self-interacting dark energy has recently garnered attention due to its implications for avoiding catastrophic scenarios such as the Big Rip \cite{Amare}. Indeed, the stability of \((\omega^{*}_{A},\omega^{*}_{B})=(-1,-1)\) in our study concur with the finding in \cite{Amare}.

However, we must emphasise that these interpretations are contingent upon the ansatz we have employed. Thus, while our exploration sheds light on the dynamic nature of an evolving dark energy, it simultaneously underscores the necessity for further investigation into the varied interactions and behaviours that could shape the universe's fate. 
 \appendix

\section{Effective equation of state}
During the isobaric period defined by the redshift interval \( z_{\text{in}} \) to \( z_{\text{out}} \), we can derive the effective equation of state for the universe as follows:

\begin{eqnarray}
\omega(z) = \alpha_{A} \omega_{A}(z) + \alpha_{B} \omega_{B}(z),
\end{eqnarray}

where \( \alpha_{A} \) and \( \alpha_{B} \) represent the relative weighting of the contributions from species \( A \) and \( B \), respectively. These weighting parameters are treated as constants in relation to the redshift during this specific period.

This formulation allows for a coherent representation of the overall equation of state, encapsulating the interactions and contributions of the competing species within the framework of our cosmological model. It follows that  

\begin{eqnarray}
\omega'(z) 
&=&3(\alpha_{A}\omega_{A}+\alpha_{B}\omega_{B})+3(\alpha_{A}\omega^2_{A}+\alpha_{B}\omega^2_{B})\nonumber\\
&+&\frac{Q_{AB}}{H} \left(\frac{\alpha_{B}\omega_{B}}{\rho_{B}}
-\frac{\alpha_{A}\omega_{A}}{\rho_{A}}\right).
\end{eqnarray}
This equation illustrates the dependence of the effective equation of state on the individual equations of state of the constituent species, the relative weighting assigned to each species, and the interactions among them. Such a formulation underscores the intricate relationships that govern the overall behaviour of the system, highlighting the significance of both the individual contributions and their interplay in shaping the effective dynamics of the equation of state. This offers us a lens into a deeper understanding of the underlying physical processes at work within the cosmological framework.

\end{document}